\documentclass[twocolumn,showpacs,preprintnumbers,amsmath,amssymb]{revtex4}
\usepackage{graphicx}% Include figure files
\usepackage{dcolumn}% Align table columns on decimal point
\usepackage{bm}% bold math
\begin{document}

\preprint{R. Palai {\it et al}.}

\title{Observation of Spin-glass-like Behavior in \sro\ Epitaxial Thin Films}

\author{\sf R. Palai$^{\ast}$, H. Huhtinen$^{\dag}$, J.F. Scott$^\ddagger$, and  R. S. Katiyar$^{\ast}$}
\affiliation{$^{\ast}$Department of Physics and Institute for
Functional Nanomaterials, University of Puerto Rico, San Juan, PR
00931-23343, USA} \affiliation{$^{\dag}$Department of Physics,
University of Turku, Turku, FIN-20014, Finland}
\affiliation{$^{\ddagger}$Department of Earth Science, University
of Cambridge, CB2 1PZ, UK}

\date{\today}% It is always \today, today,

\renewcommand{\baselinestretch}{} % ... as the instructions say
\newcommand{\bfo}{BiFeO$_{3}$}% for BiFeO3
\newcommand{\sro}{SrRuO$_{3}$}% for SrRuO3
\newcommand{\tn}{$T_{\rm N}$}% for Neel temperature
\newcommand{\pr}{$P_{\rm r}$}% for electrical polarization
\newcommand{\mr}{$M_{\rm r}$}% magnetisation
\newcommand{\tc}{$T_{\rm c}$}%Curie Temperature
\newcommand{\jc}{$J_{\rm c}$}% Current density
\newcommand{\rs}{$R^{\rm 2}$}% for R square
\newcommand{\dc}{$^{\circ}$C}% for degree centigrade
\newcommand{\rr}{$r_{\rm R}$}% for resistance ratio
\newcommand{\hc}{$H_{\rm c}$}% for resistance ratio
\newcommand{\ms}{$M_{\rm s}$}% for resistance ratio
\newcommand{\tirr}{$T_{\rm irr}$}% for resistance ratio
\newcommand{\tf}{$T_{\rm F}$}% for resistance ratio
\newcommand{\mirr}{$M_{\rm irr}$}% for resistance ratio
\newcommand{\mfc}{$M_{\rm FC}$}% for resistance ratio
\newcommand{\mzfc}{$M_{\rm ZFC}$}% for resistance ratio
\newcommand{\tgh}{$T_{\rm g}$($H$)}% for resistance ratio
\newcommand{\tg}{$T_{\rm g}$}% for resistance ratio

\begin{abstract}

We report the observation of spin-glass-like behavior and strong
magnetic anisotropy in extremely smooth ($\sim$~1-3~\AA\
roughness) epitaxial (110) and (010) SrRuO$_{3}$ thin films. The
easy axis of magnetization is always perpendicular to the plane of
the film (unidirectional) irrespective of crystallographic
orientation. An attempt has been made to understand the nature and
origin of spin-glass behavior, which fits well with Heisenberg
model.

\end{abstract}

\maketitle
\section{INTRODUCTION}

Integration of functional materials (oxides of ferroelectrics and
multiferroics) into silicon technology is of great technological
and scientific interest. The current interest in functional oxides
is largely based on engineered epitaxial thin films because of
their superior properties compared to the bulk and polycrystalline
thin films and their technological applications in dynamic random
access memories, magnetic recording, spintronics, and sensors
\cite{ramesh:sc02, scott:sc07, scott:nm07}. Most of these
applications require bottom and top electrodes to exploit the
electronic properties of the functional materials.

\sro\ (SRO) has been found to be very useful for electrodes and
junctions in microelectronic devices because of its good
electrical and thermal conductivity, better surface stability, and
high resistance to chemical corrosion, which could minimize
interface electrochemical reactions, charge injection in oxide,
and other detrimental processes \cite{ahn:apl97,lee:apl04} thus
improving retention, fatigue resistance, and imprint. It also has
good work function to produce the required large Schottky barrier
on most ferroelectric oxide capacitors \cite{hartmann:apa}. Growth
of an atomically flat epitaxial SRO film is required for a smooth
and stable interface which is essential for the growth of
subsequent layers, as high imperfections and roughness in the base
layer can induce defects in the upper layers, which can
irreversibly destroy the material properties. However, recent
studies of epitaxial thin films
\cite{klein:apl95,gan:apl98,kang:apl05} suggest that SRO may have
novel magnetostructural properties in ultra thin film form. By
tuning film thickness and in-plain strain very different
properties may emerge as compared with bulk. It has been found
that  thin films of SRO show uniaxial magnetic anisotropy
 \cite{klein:apl95, gan:apl98} instead of biaxial anisotropy observed in bulk
\cite{kanabayasi:jpsj76, cao:prb97}.

The bulk SRO exhibits an orthorhombic crystal structure
($a$~=~5.570 \AA, $b$~=~5.530 \AA, and $c$~=~7.856 \AA)
\cite{jones:csc89} and several useful properties, such as
extraordinary Hall effect \cite{ahn:apl97}, strong
magnetocrystalline anisotropy \cite{klein:apl95}, itinerant
ferromagnetism \cite{callaghan:ic66, bouchard:mrs72}, and
spin-glass behavior \cite{reich:jmmm99}. Spin-glass materials are
currently frontier field of research and the most complex kind of
condensed state of matter encountered so far in solid state
physics. Despite of the enormous importance of spin-glass models
in neural networks \cite{amit:pra85}, our knowledge of the
underlying mechanistic processes involved in is extremely limited.
Some of the typical features of spin glass are: spin freezing
(very slow time relaxation of magnetization); a cusp in the
temperature dependence of magnetization; irreversible behavior of
magnetization below the freezing temperature; remanence; magnetic
hysteresis \cite{binder:rmp, fischer:book1991}. Although
spin-glass-like behavior has been reported in bulk SRO, to our
knowledge, the behavior is not well understood and there was no
such report in thin films. In this letter, we report the
observation, interpretation, and possible origin of
spin-glass-like behavior in very smooth epitaxial (110) and (010)
SRO thin films and observation of spontaneous alignment of domains
in (010) thin films.
\begin{figure}[h!]
\begin{center}
\includegraphics [width=0.4\textwidth,clip]{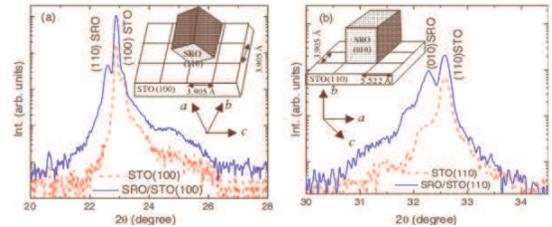}
\caption{\sf XRD patterns of SRO (110) (a) and (010) (b) films
grown on (100) and (110) STO substrates, respectively. XRD pattern
of STO is given for comparison. The insets depict the schematic of
the growth orientations of SRO films on (100) and (110) STO
substrates.} \label{fig:xrd}
\end{center}
\end{figure}
\section{EXPERIMENTAL DETAILS}
We investigated SRO thin films of 25~nm thick grown on (100) and
(110) SrTiO$_{3}$ (STO) substrates of area (5~mm)$^{2}$ by pulsed
laser deposition (PLD). The growth parameters were as follow:
substrate temperature of 750~\dc, oxygen partial pressure of
100~mTorr, laser energy density of 2.0~Jcm$^{-2}$ at a pulse rate
of 10~Hz. X-ray diffraction (XRD) was used to investigate the
orientation and crystallinity of films. Microstructure and growth
mechanism of the films were studied using atomic force microscopy
(AFM). Magnetic measurement were carried out using a
superconducting quantum interference device (SQUID) magnetometer.
Before the magnetic measurement the silver paint was removed from
the back of the substrate to eliminate spurious magnetic signal.

\begin{figure}[h!]
\begin{center}
\includegraphics [width=0.45\textwidth,clip]{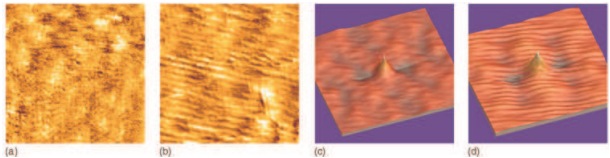}
\caption{\sf AFM images of SRO films on STO substrates; (a) [110]
oriented SRO film on [100] STO substrate; (b) [010] oriented SRO
film on [110] STO substrate; (c) and (d) are the 3D
autocorrelation images of the surfaces (a) and (b), respectively.
The scan area was 2~x~2~$\mu$m$^{2}$.} \label{fig:afm}
\end{center}
\end{figure}
\section{RESULTS AND DISCUSSION}
The XRD patterns (Fig.~\ref{fig:xrd}) of 25 nm thick SRO films on
(100) and (110) STO (cubic, $a$~=~3.905~\AA) substrates show [110]
and [010] orientation, respectively. The insets in
Fig.\ref{fig:xrd} depict the schematic of growth orientation of
SRO films on (100) and (110) STO substrates. The out-of-plane
lattice parameter of [110] and [010] oriented SRO films was found
to be $d_{110}$~=~3.932~\AA, and $d_{010}$~=~5.543~\AA,
respectively, which is slightly larger than the corresponding bulk
value of $d_{110}$~=~3.924~\AA, and $d_{110}$~=~5.530~\AA. This
clearly implies that films have out-of-the-plane tensile strain.
The in-plane lattice strain between STO$_{100}$ and SRO$_{110}$
and STO$_{110}$ and SRO$_{010}$ was calculated and found to be
0.5\% compressive  and this should result out-of-the-plane lattice
parameter of 3.939~\AA\ in [110] and 5.557~\AA\ in [010]
orientated films, which agrees with our observed values as strain
gradually relaxes with increasing thickness.

Fig.~\ref{fig:afm} shows the AFM images of SRO films on
2~x~2~$\mu$m$^{2}$ area. The surface morphology of both the films
did not now show any 3D-like island or spiral-like growth, but
rather 2D-like layer-by-layer growth. However, a spontaneous
alignment of the grains (magnetic domains) has been observed in
[010] oriented film. The films were atomically smooth and the
surface roughness ($Z$$_{\rm rms}$) was found to be 1.3 and
2.1~\AA\ in [110] and [010] oriented films on 2~x~2~$\mu$m$^{2}$
area, respectively, which is close to the AFM resolution.
Functions, such as  2D Fourier transform or autocorrelation
function can be used to quantify the aspect of the texture and
lateral directionality of the surface topography. The asymmetry in
the autocorrelation function quantifies the directionality of the
features \cite{bonnell:book2001}. A scanning probe microscopy
software \cite{wsxm:sw2005} has been used for analyzing the
autocorrelation function of the AFM images. Figs.~\ref{fig:afm}(c)
and (d) are the 3D autocorrelation image of AFM surfaces (a) and
(b), respectively. The symmetric nature of autocorrelation images
shows excellent lateral directionality of the SRO films.

\begin{figure}[h!]
\begin{center}
\includegraphics [width=0.4\textwidth,clip]{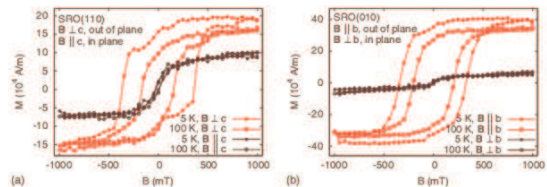}
\caption{\sf Temperature dependence of hysteresis (M-B) loops of
SRO films on STO substrates at different orientations; (a) (110)
SRO film; (b)(010) SRO film.} \label{fig:hys}
\end{center}
\end{figure}

Fig.~\ref{fig:hys} shows temperature dependence of hysteresis
loops of (110) and (010) SRO films with applied field ({\it B})
perpendicular ($\perp$) and parallel ($\parallel$) to the plane of
the film. The [110] oriented film showed maximum saturation
magnetization (\ms) of about 19 $\times$ 10$^{4}$ $A$/$m$ (190
emu/cc), while [010] oriented film showed two times higher
magnetization than [110] oriented film. The possible explanation
could be the presence of spontaneous alignment of the magnetic
domains in [010] oriented film. The anomalies observed in the
hysteresis loop at 5~K (less acute in the case of (010) film)
between two regions of opposite field could be the Barkhausen
jumps \cite{stohr:book06}. These jumps are generally caused by the
irreversible motion of the domain walls between the two regions of
opposite magnetizating forces. As evident from Fig.~\ref{fig:hys},
there is only one easy axis of magnetization and it is always
out-of-plane (perpendicular to the plane of the film) and
perpendicular to the {\it c}-axis, contrary to the earlier
observation of easy axis along the $c$-axis on SRO films on STO
(100) substrates \cite{Izumi:jjap97}, but in agreement with
\cite{klein:apl95,gan:jap99}. The observed strong magnetic
anisotropy could be the manifestation of spin-orbital coupling of
ruthenium atoms or possibly due to the strong pinning of the
domains perpendicular to the film. Materials with the easy axis of
magnetization perpendicular to the surface have considerable
importance in realizing the next generation perpendicular magnetic
recording (PMR) system. Almost all the commercial recording
systems available in the market use magnetic media with
magnetization in the plane of film, known as longitudinal magnetic
recording (LMR) and is limited by the superparamagnetic effect.
The storage densities as high as 1Tbit in$^{-2}$ could be achieved
with the PMR system \cite{wang:nm05,vopsaroiu:jap08}, where the
superparamagnetic effect is less acute.
\begin{figure}[h!]
\begin{center}
\includegraphics [width=0.45\textwidth,clip]{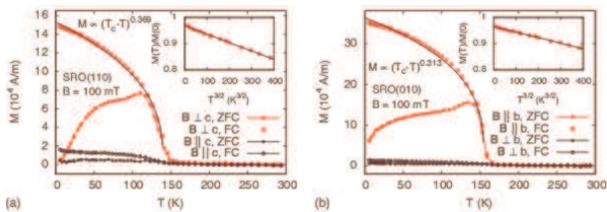}
\caption{\sf (a) ZFC and FC magnetizations of SRO films as a
function of temperature at different orientations; (a) (110)SRO
films; (b) (010) SRO film. The solid line on FC magnetizations is
the curve fitted with  M $\propto$ (\tc-$T$)$^{\beta}$. The insets
show the reduced magnetization $M$($T$)/$M$(0) {\it vs}.
$T$$^{3/2}$ in the low-temperature region. The straight lines are
linear fit to the Bloch's Law.} \label{fig:mt}
\end{center}
\end{figure}

Temperature dependence of zero-field cooled (ZFC) and field cooled
(FC) magnetizations of (110) and (010) SRO films recorded in
different orientations with 100~mT are shown in Fig.\ref{fig:mt}.
As can be seen, an unidirectional (always out-of-plane) anisotropy
has been observed irrespective of the orientations. The
magnetization also measured field-in-plane perpendicular to the
$c$-axis in SRO(110) films, but was found to be magnetically hard.
The transition temperature was found to be 150 and 160~K for [110]
and [010] orientated films, respectively. The out-of-plane FC
spontaneous magnetization below \tc\ follows the scaling law,
$M$~=~C(\tc-$T$)$^{\beta}$, with critical exponent $\beta$~=~0.369
($\pm$~0.006)  and Curie constant, C~=~2.46 ($\pm$~0.06) $A$/$m$
for [110] oriented film indicating 3D Heisenberg-like ferromagnet
(for which theory gives $\beta$~=~0.367) \cite{blundell:book01};
whereas $\beta$~=~0.313 ($\pm$~0.007), and C~=~7.45 ($\pm$~0.25)
$A$/$m$ were obtained for [010] oriented film implying Ising
(3D)-like ferromagnet (theoretical value of $\beta$~=~0.326)
\cite{blundell:book01}. However, there are some discrepancies in
the literature. The exponent $\beta$~=~0.43 \cite{Wang:prb04} and
0.325 \cite{klein:prl96} was obtained for [110] oriented 100 nm
thick films, while $\beta$~=~0.5  was obtained for single-crystal
\cite{kim:prb03} and interpreted as mean-field behavior (in our
opinion this results from strain, which is always unscreened and
hence long-range). The difference in exponent value in our case
could be related to different domain structure, orientation or
strain effect. A more detailed study of these behaviors we
describe below. The insets in Figs.\ref{fig:mt}(a) and (b) show
plots of reduced magnetization $M$($T$)/$M$(0) {\it vs}.
$T$$^{3/2}$ in the low temperature region, where $M$(0)\ is the
magnetization at 0~K, with the linear fit to the data. As evident,
the magnetization behavior well describes the Bloch's Law
$M$($T$)/$M$(0)~=~1~-~$A$ $T$$^{3/2}$ (where $A$ is the spin wave
parameter) implying the dominance of spin-wave excitation on
magnetization as expected for a Heisenberg ferromagnet
\cite{mueller:epj99}. In itinerant ferromagnets, the low
temperature magnetization is further suppressed by a term
$T$$^{2}$, which is due to Stoner excitations of magnetic
electrons. The excellent fit to $T$$^{3/2}$ law is clearly
observed in both the films even without the small $T$$^{2}$
correction, which implies the suppression of Stoner excitations.
The fact that FC magnetization does not saturate at low
temperature implies short-range spin ordering in SRO like spin
glasses La$_{0.7-x}$Nd$_{x}$Pb$_{0.3}$MnO$_{3}$ (x~=~0.5 and 0.7),
which also follow Bloch's $T$$^{3/2}$ law \cite{young:jjap01}. The
exchange interaction ($J$) between two neighboring Ru$^{4+}$ ions
was calculated using spin-wave parameter,
$A$~=~(0.0587/$S$)($k$$_{B}$/2$J$$S$)$^{3/2}$, where $S$ is the
total spin of Ru$^{4+}$ and $k$$_{B}$ is Boltzmann's constant, and
found to be 16.1~$k$$_{B}$~K and 21.37 $k$$_{B}$~K for [110] and
[010] oriented films, respectively. The larger exchange energy for
[010] oriented films is in agreement with the higher \tc\
observed. In comparison, $J$ values of 14.41~$k$$_{B}$~K and
20.57~$k$$_{B}$~K have been reported for 100~nm thick SRO films
deposited by sputtering \cite{Wang:prb04}.
\begin{figure}[h!]
\begin{center}
\includegraphics [width=0.4\textwidth,clip]{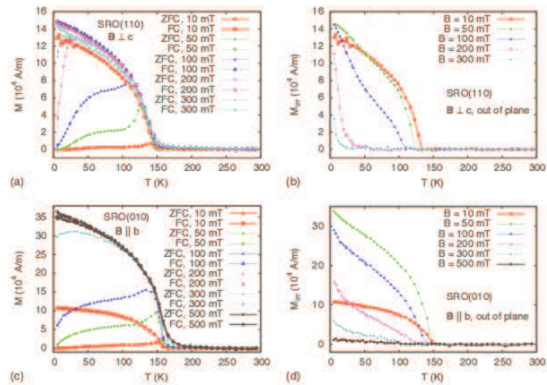}
\caption{\sf Temperature dependence of out-of-plane ZFC and FC
magnetizations of (110) (a) and (010) (c) SRO films with their
corresponding irreversibility magnetization \mirr\ vs. $T$ (b)
(110) and (d) (010) SRO thin films.} \label{fig:mtall}
\end{center}
\end{figure}

Figs.~\ref{fig:mtall} (a) and (c) show the temperature dependence
of ZFC and FC magnetization at different fields applied
out-of-plane (perpendicular to the plane of the film) for [110]
and [010] oriented films, respectively. A significant difference
has been observed in ZFC and FC magnetization. A closer
observation of out-of-plane ZFC magnetization reveals three
characteristic features: a critical temperature \tc, the onset of
nonzero {\it M}, an irreversibility temperature \tg, where the ZFC
and FC branches coalesce, and a pronounced cusp, which varies with
field and gradually smoothen at higher fields. According to
Edwards and Anderson \cite{edwards:jpf75} the latter one is
because of the interaction of the spins dissolved in the matrix,
as a result there is {\it no} mean ferro or antiferromagnetism,
but there will be a ground state with the spins aligned in
definite directions. It is interesting to note that the
out-of-plane FC magnetization with 10~mT for [010] oriented film
is less compared to the [110] orientated film. This can understood
due to spontaneous alignment of the domains, the field of 10~mT is
not adequate to align all the domains along the direction of the
field.

The irreversible magnetization \mirr~(= \mfc~-~\mzfc) as a
function of temperature for different fields is shown in
Figs.~\ref{fig:mtall}(b) and (d) for [110] and [010] oriented
films, respectively. The point at which \mirr\ becomes nonzero
defines a spin-glass transition temperature, \tg\ at the working
field and the characteristic measuring time. The field dependence
of \tgh, which increases with field, is the most important
characteristic of spin glasses due to competing interactions
because of frozen disorders and magnetic frustrations
\cite{binder:rmp,fischer:book1991,gruyter:prl05}. A spin glass
order parameter can be estimated from the field dependence of
\tgh\ that vanishes roughly linearly with temperature at freezing
temperature (\tf) \cite{binder:rmp}. Note that a spin glass canot
be described by a single order parameter, but rather requires many
of them due to the existence of many phases
\cite{fischer:book1991}.

The exact nature of frozen disorder and frustration is not quite
clear, however, possibly \emph{spin canting} at low temperature
might have produced \emph{finite} spin clusters (composed of a set
of non-collinear ferromagnetically or antiferromagnetically
coupled spins), which are embedded in the \emph{infinite} 3D
ferromagnetic (FM) matrix \cite{fischer:book1991, kaul:cs05}.

\begin{figure}[h!]
\begin{center}
\includegraphics [width=0.4\textwidth,clip]{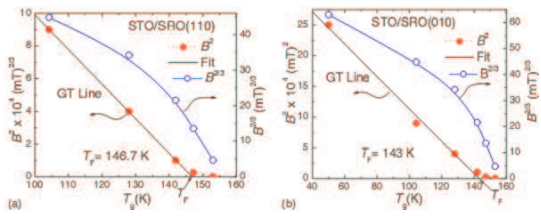}
\caption{\sf Field dependence of \tg raised to the square and 2/3
 power for (110) (a) and (010) (b) SRO films. The solid lines are
 the fitted to data points with Eq.~\ref{equn2}.}
\label{fig:bt}
\end{center}
\end{figure}

In order to understand the frozen state and freezing transition of
SRO thin films, the behavior of magnetic field has been analyzed
in the field-temperature plane. The existence of critical lines
(Fig.\ref{fig:bt}) can be explained by mean field theory (MFT) in
the framework of replica-symmetry breaking \cite{binder:rmp,
fischer:book1991, gruyter:prl05}. The equations for the transition
lines have been predicted by de Almeida and Thouless (called AT
line) for Ising spin glass with infinite-range random interactions
and by Gabay and Toulouse (called GT line) for the Heisenberg spin
glass \cite{binder:rmp, fischer:book1991,
gruyter:prl05,lefloch:physicab94}.

The so called AT line is usually defined as $H$(\tg) and behaves
near the freezing temperature as:
\begin{equation}
\left(1- \frac{T_{\rm g}(H)}{T_{F}}\right) ^{3} =
\frac{3}{4}h^{2},~ {\rm with}~ h = \frac{\mu H}{k_{B}T_{F}},
 \label{equn1}
\end{equation}
where \tf\ is the zero-field spin-glass freezing temperature. The
GT line is defined as:

\begin{equation}
1- \frac{T_{\rm g}(H)}{T_{F}} = \frac{m^{2}+4m +2}{4(m + 2)^{2}}
h^{2},
 \label{equn2}
\end{equation}
where $m$ is the total number of the components (with $m$~-~1
transverse components) of the spin glass. At this line only the
{\it transverse} components of the spins should be freezing-in at
the low temperature, while the freezing-in of longitudinal
components should occur at the cross-over, which is similar to the
AT line. The critical lines are defined by a dynamical instability
and onset of broken ergodicity, as manifested by irreversible
effect due to the existence of a large number of degenerate
thermodynamic states with the same microscopic properties but with
different microscopic configurations \cite{binder:rmp}. The
existence of critical line is one of the important fingerprints of
spin glass \cite{binder:rmp, fischer:book1991,
gruyter:prl05,lefloch:physicab94}. In Fig.~\ref{fig:bt}, the field
dependence of \tg\ was raised to square and 2/3 power in order to
check the critical lines. Note that in this system it is difficult
to clearly differentiate between Ising and Heisenberg spin
glasses. However, as can be seen, the experimental data shows a
better fitting with Eq.~\ref{equn2} with GT line for $H$ $>$
100~mT implying Heisenberg-like spin-glass and the \tf\ was found
to be $\approx$ 146.7 ($\pm$~0.19) and 143 ($\pm$~3)~K for [110]
and [010] oriented films, respectively. It is intriguing to note
that the FC out-of-plane magnetization of [010] films at low field
(100 mT) follows Ising-like ferromagnet (3D), but the overall
behavior in $H$-$T$ phase space agrees well with Heisenberg model.
This can be understood as spin glasses have many metastable spin
configurations and dynamics on many timescales and a complicated
behavior is expected when spin glass coexists with FM orderings,
called \emph{magnetized spin glass} \cite{binder:rmp}.
\begin{figure}[h!]
\begin{center}
\includegraphics [width=0.3\textwidth,clip]{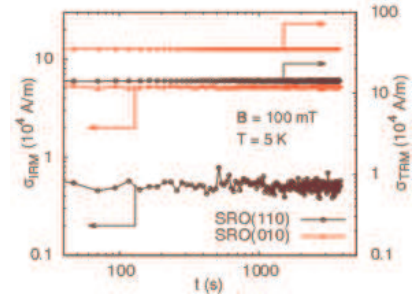}
\caption{\sf Time relaxation of isothermal remanent magnetization
($\sigma$$_{IRM}$) and thermoremanent magnetization
($\sigma$$_{TRM}$ ) of SRO thin films. The "open circle"
represents for [110] films, while "solid circle" for [010] films.}
\label{fig:mtime}
\end{center}
\end{figure}
\begin{figure}[h!]
\begin{center}
\includegraphics [width=0.3\textwidth,clip]{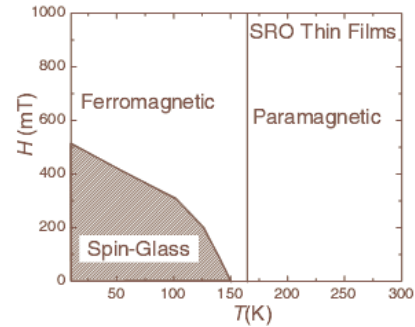}
\caption{\sf Schematic of field-temperature ($H$-$T$) phase
diagram of SRO thin films.} \label{fig:phase}
\end{center}
\end{figure}

In order to understand the characteristic excitation and
relaxation time, we investigated both the isothermal remanent
magnetization ($\sigma$$_{IRM}$) and thermoremanent magnetization
($\sigma$$_{TRM}$) of [110] and [010] oriented SRO films at 100~K
(not shown) and 5~K [Fig.\ref{fig:mtime}]. The $\sigma$$_{IRM}$
was measured cooling the samples in zero field (ZFC) to the
desired temperature to be studied; then a field of (100 mT) was
applied for 10 mins and switched off again, and the time
relaxation was followed. To obtain the $\sigma$$_{TRM}$, on the
other hand, the sample was field cooled (100 mT) at some initial
temperature (RT) above \tf\ and then the system was slowly cooled
down in a constant field to the desired temperature (100K or 5K),
at which the field was switched off and the time relaxation was
followed. As can be seen from Fig.\ref{fig:mtime}, no exponential
time-decay, but rather a very {\it slow relaxation}, has been
observed over macroscopic time scale indicating a large misfit at
the domain walls causing extremely slow domain growth
\cite{binder:rmp}. An extremely slow relaxation of magnetization
with time below \tf\ is particularly an important and interesting
salient feature of spin-glasses \cite{binder:rmp,
fischer:book1991}. This type of slow relation was reported in
other FM spin glass materials {\it i.e.}, La$_{2}$CoMnO$_{6}$
\cite{wang:scst02} and AuFe (Heisenberg-like FM
\cite{campbell:prl83}) alloy \cite{binder:rmp, fischer:book1991}.

A schematic of the phase diagram [Fig.~\ref{fig:phase}] of SRO
thin films was drawn from all measurements. We found that the spin
glass behavior is confined within $\sim$~1~T and $\sim$~150~K.

In \sro\, Ru ions, known to be the only site of magnetic moment,
are arranged in a strictly periodic order, which is unfavorable
for a spin glass state. SRO with tolerance factor (Goldschmidt) of
0.994 would not be expected to be distorted, but, the polarized
neutron-scattering experiments on single crystal showed the strong
hybridization of Ru(4d)-O(2p) orbitals, which results in 10\%
 of ordered magnetic moment associated with oxygen site \cite{nagler:baps97}; approximately
30\% has been theoretically calculated \cite{mazin:prb97}. We
believe that this distribution of magnetic moments between the
Ru-sites and the O-sites might have created the frustration and
the randomness necessary for the spin glass. Recently, Zayak {\it
et al.} \cite{Zayak:prb06} have predicted  that SRO also has weak
A- and C-type antiferromagnetic (AFM) spin configurations along
with most stable FM configuration. Our conjecture is that some AFM
spin clusters (at Sr-site) could have embedded into the FM matrix
(at Ru-site) causing the randomness necessary for the origin of
spin glass behavior. Epitaxial strain and small oxygen vacancy
could trigger the change of the spin configurations and the shape
of the octahedra in SRO thin films.

The spin glass behavior can also be understood from the
discrepancy between the calculated magnetic moment of
2.82$\mu_{\rm B}$ (for S~=~1 on the spin-only formula) and the
measured magnetic moment from the saturation magnetization
corresponds to the Curie constant of 2.46 $A$/$m$ for [010]
oriented film, leading to $\mu$~=~3.19 $\mu_{\rm B}$/formula unit
(f.u.). The part of the moment which is being frozen out below \tf
could be related to {\it magnetic domains}.

\section{CONCLUSION}
In conclusion, spin glass-like behavior was observed in high
quality [110] and [010] oriented SRO films grown on STO substrates
by PLD. AFM images of [010] orientated films showed spontaneous
alignment of domains. An {\it unidirectional} anisotropy was
observed; easy axis of magnetization was always perpendicular to
the surface of the thin films irrespective of film orientation,
which has immense importance in the next generation of magnetic
recording media.

\section*{ACKNOWLEDGEMENT}
The financial support from DEPSCoR W911NF-06-0030 and
W911NF-05-1-0340 grants is gratefully acknowledged. RP thanks
Institute for Functional Nanomaterials (IFN), University of Puerto
Rico for financial support.

%\newpage
\section*{}


\begin{thebibliography}{29}

\bibitem{ramesh:sc02} R. Ramesh and D.G. Schlom, Science {\bf 296}, 1975 (2002).

\bibitem{scott:nm07} J. F. Scott, Nature Materials {\bf
6}, 256 (2007).

\bibitem{scott:sc07} J. F. Scott, Science {\bf 315}, 954 (2007).


\bibitem{ahn:apl97} C.H. Ahn, R.H. Hammond, T.H. Geballe, M.R. Beasley, J.M. Triscone, M. Decroux, $\O$. Fisher, L.
Antognazza, and K. Char, Appl. Phys. Lett. {\bf 70}, 206 (1997).

\bibitem{lee:apl04} H.N.Lee, H.M. Christen, M.F. Chisholm, C.M. Rouleau, and D.H. Lowndes,
Appl.Phys. Lett. {\bf 84}, 4107 (2004).

\bibitem{hartmann:apa} A.J. Hartmann, M. Neilson, R.N. Lamb, K. Watanabe, and J.F. Scott,
Appl. Phys. A  {\bf 70}, 239 (2000).


\bibitem{klein:apl95} L. Klein, J.S. Dodge, T.H. Geballe, A. Kapitulnik, A.F. Marshall, L. Antognazza, and K. Char,
Appl. Phys. Lett. {\bf 66}, 2427 (1995).


\bibitem{gan:apl98} Q. Gan, R.A. Rao, C.B. Eom, J.L. Garrett, and M. Lee, Appl. Phys. Lett. {\bf 72}, 978 (1998).

\bibitem{kang:apl05} B.S. Kang, J-S. Lee, L. Stan, L. Civale, R.F. DePaula, P. N. Arendt, and Q.X. Jia,
Appl. Phys. Lett. {\bf 86}, 072511 (2005).

\bibitem{kanabayasi:jpsj76} A. Kanabyasi, J. Phys. Soc. Japan {\bf 41}, 1876
(1976).

\bibitem{cao:prb97} G. Cao, S. McCall, M. Shepard, J.E. Crow, and
R. P. Guertin, Phys. Rev. B {\bf 56}, 321 (1997).

\bibitem{jones:csc89} C.W.Jones, P.D. Battle, P. Lightfoot, W.T.A. Harrison, Acta.
Crystallogr:C Cryst. Struct. Commun. {\bf 45}, 365 (1989).

\bibitem{callaghan:ic66} A. Callaghan, C.W. Moller, and R. Ward, Inorg.Chem. {\bf 5},  1572
(1966).
\bibitem{bouchard:mrs72} R.J. Bouchard and J.L. Gillson, Mater. Res. Bull. {\bf 7},  893
(1972).

\bibitem{reich:jmmm99} S. Reich, Y. Tsabba, and G. Cao, J. Magn.
Magn. Mater. {\bf 202}, 119 (1999).

\bibitem{amit:pra85} D.J. Amit, H. Gufreund, and H. Sompolinsky, Phys. Rev. A {\bf 32}, 1007 (1985)
and references therein.

\bibitem{binder:rmp} K. Binder and A.P. Young, Rev. Mod. Phys. {\bf 58}, 801 (1986).

\bibitem{fischer:book1991} K.H. Fischer and J.A. Hertz,  {\it Spin
glasses}, Cambridge University Press. (1993).


\bibitem{bonnell:book2001} D. Bonnell (Ed) {\it Scanning probe
microscopy and spectroscopy: theory, techniques, and
applictaions}, 2nd ed, Wiley-VCH, Inc.,(2001).

\bibitem{wsxm:sw2005} Scanning probe microscopy software (WSxM), Nanotec. Electronica S. L. (2005).

\bibitem{stohr:book06} J. St\"{o}hr and H.C. Siegmann {\it Magnetism From Fundamental to Nanoscale Dynamics
}, Springer, Berlin, p. 516, (2006).
\bibitem{Izumi:jjap97} M. Izumi, K. Nakazawa, Y. Bando, Y. Yonedo, and H. Terauchi,
Jpn. J. Appl. Phys. {\bf 66}, 3893 (1997).

\bibitem{gan:jap99} Q. Gan, R.A. Rao, C.B. Eom, L. Wu, and F. Tsui, J. Appl. Phys. {\bf 85}, 5297 (1999).

\bibitem{wang:nm05} J. Wang, Nature Materials {\bf 4}, 191 (2005).
\bibitem{vopsaroiu:jap08} M. Vopsaroiu, J. Blackburn, A. Muniz-Piniella, and M.G. Cain,
J. Appl. Phys. {\bf 103}, 07F506 (2008).

\bibitem{blundell:book01} S. Blundell, {\it Magnetism in Condensed Matter}, Oxford University Press, p. 119, (2001).

\bibitem{Wang:prb04} L.M. Wang, H.E. Horng, and H.C. Yang, Phys. Rev. B {\bf 70}, 014433 (2004).
\bibitem{klein:prl96} L. Klein, J.S. Dodge, C.H. Ahn, G.J. Snyder, T.H. Geballe, M.R. Beasley, and A. Kapitulnik,
Phys. Rev. Lett. {\bf 77}, 2774 (1996).

\bibitem{kim:prb03}  D. Kim, B.L. Zink, F. Hellman, S. McCall, G. Cao, and J.E. Crow,
Phys. Rev. B {\bf 67}, 100406 (2003).

\bibitem{mueller:epj99} M. Mueller, U. K\"{o}bler, and K. Fisher, Euro. Phys. J. B {\bf 8}, 207 (1999).

\bibitem{young:jjap01} S.L. Young, H.Z. Young, H.Z. Chen, L. Horng, J.B. Shi, and Y.C. Chen,
Jpn. J. Appl. Phys. {\bf 40}, 4878 (2001).


\bibitem{edwards:jpf75} S.F. Edwards and P.M. Anderson, J. Phys. F {\bf 5}, 965 (1975).

\bibitem{kaul:cs05} S.N. Kaul, Current. Sci. {\bf 88}, 78 (2005).

\bibitem{gruyter:prl05} M. Gruyter, Phys. Rev. Lett. {\bf 5}, 965 (1975).

\bibitem{lefloch:physicab94} F. Lefloch, J. Hammann, M. Ocio, and E. vincent, Physica B {\bf 203}, 63 (1994).


\bibitem{wang:scst02} X.L. Wang, M. James, J. Horvat, and S.X. Dou, Suppercond. Sci. Technol. {\bf 15}, 427 (1994).


\bibitem{campbell:prl83} I.A. Campbell, S. Senoussi, F. Varret, J. Teillet, and A.
Hamzi\'{c}, Phys. Rev. Lett. {\bf 50}, 1615 (1983).


\bibitem{nagler:baps97} S. Nagler and B.C. Chakoumakos, Bull. Am. Phys.
Soc. {\bf 42}, 551 (1997).

\bibitem{mazin:prb97} I.I. Mazin and D.J. Singh, Phys. Rev. B {\bf 56}, 2556 (1997).

\bibitem{Zayak:prb06} A.T. Zayak, X. Huang, J.B. Neaton, and K.M. Rabe, Phys. Rev. B {\bf 74}, 094104 (2006).


\end{thebibliography}
\end{document}